\documentclass[12pt,letterpaper]{article}

\usepackage[includeheadfoot,
            marginratio={1:1,2:3}, 
            width=412pt, 
            height=688pt,]{geometry}

\usepackage{amsmath}
\usepackage{amsfonts}
\usepackage{amssymb}
\usepackage{graphicx}
\usepackage{amsthm}
\usepackage{simplewick}

\newcommand{\eq}[1]{\begin{equation}
                     \begin{split} #1 \end{split}
                     \end{equation}}

\newcommand{\ov}{\overline}

\allowdisplaybreaks[2]
\numberwithin{equation}{section}

\begin{document}

\vspace*{-1.5cm}
\begin{flushright}
  {\small
  MPP-2017-126\\
  }
\end{flushright}

\vspace{1.5cm}
\begin{center}
{\LARGE
On the Structure of Quantum L$_\infty$ algebras\\[0.3cm]

}
\vspace{0.4cm}

\end{center}

\vspace{0.35cm}
\begin{center}
  Ralph Blumenhagen, Michael Fuchs, Matthias Traube
\end{center}

\vspace{0.1cm}
\begin{center} 
\emph{Max-Planck-Institut f\"ur Physik (Werner-Heisenberg-Institut), \\ 
   F\"ohringer Ring 6,  80805 M\"unchen, Germany } \\[0.1cm] 

\vspace{0.4cm} 
 
\end{center} 

\vspace{1cm}


\begin{abstract}
It is believed that any classical gauge symmetry gives rise to an
L$_\infty$ algebra. Based on the recently realized relation between 
classical ${\cal W}$ algebras and L$_\infty$ algebras,   we analyze how this generalizes to
the quantum case.   Guided by the existence of quantum ${\cal W}$ algebras, we provide a physically well motivated definition of quantum 
L$_\infty$ algebras describing the  consistency of global symmetries in quantum field theories.
In this case  we are restricted to only two  non-trivial
 graded vector spaces $X_0$ and $X_{-1}$ containing the symmetry variations and the symmetry generators.
This quantum L$_\infty$ algebra structure
is explicitly exemplified for the quantum ${\cal W}_3$ algebra.
The natural  quantum product between fields is the  normal ordered one
so that, due to contractions between quantum fields, the
higher L$_\infty$ relations receive off-diagonal quantum corrections.
Curiously, these are  not present in the loop L$_\infty$ algebra of closed string
field theory.
\end{abstract}


\clearpage


\section{Introduction}

Derived from closed string field theory \cite{Zwiebach:1992ie}, the structure of L$_\infty$ algebras were suggested to underly all classical perturbative gauge symmetries and their dynamics. For the first time, they actually
appeared in the context of higher spin gauge theories \cite{Berends:1984rq} and 
were also discussed in the mathematics literature (see e.g. \cite{Lada:1992wc,Roytenberg:1998vn,Fulp:2002kk,Deser:2016qkw}). 
Motivated by the study of field theory truncations of string field
theory \cite{Sen:2016qap}, the authors of \cite{Hohm:2017pnh} argued  that the symmetry and the action of any consistent perturbative gauge symmetry is controlled by an L$_\infty$ algebra.
For Chern-Simons and Yang-Mills gauge theories as well as for double
field theory the symmetries and equations of motion could be expressed
in terms of an  L$_\infty$ structure.

{ Based on the higher spin AdS$_3$-CFT$_2$ duality}, a large set  of explicit non-trivial L$_\infty$ algebras 
were identified recently \cite{Blumenhagen:2017ogh} by showing that the well understood
class of   classical
${\cal W}$ algebras can also be rewritten in terms of  higher products
satisfying the relations of  L$_\infty$ algebras. Recall that ${\cal W}$ algebras 
appear as extended chiral symmetry algebras of two-dimensional
conformal field theories (CFTs)(see \cite{Bouwknegt:1992wg} for a
review) {and are actually not describing gauge symmetries but  infinitely many
global symmetries.}
These examples are
special in the sense that only two graded vector spaces were
non-trivial, $X_0$ contains the symmetry parameters and $X_{-1}$ the
generators of the ${\cal W}$ algebra. The special feature 
of ${\cal W}$ algebras, namely that the Poisson bracket between
the generators closes only non-linearly, implied non-trivial higher
products, corresponding e.g. to field dependent symmetry  parameters.

In \cite{Blumenhagen:2017ogh} this correspondence was restricted to the classical case,
for which the product of fields is just the point-wise product of
holomorphic functions. However, from CFT it is well known that these
classical ${\cal W}$ algebras appear as the classical $\hbar\to 0$
limit of quantum ${\cal W}$ algebras. Here one is dealing with chiral 
quantum fields, whose product involves a normal ordering prescription.
In addition, the field content of the algebra itself and their
structure constants receive $\hbar$ corrections.

It is an interesting question, how the L$_\infty$  structure generalizes to the
quantum case. In the context of string  field theory, this was already
analyzed in \cite{Zwiebach:1992ie} and further elucidated in the mathematical
context in \cite{Markl:1997bj}.
In this paper we generalize the analysis of
\cite{Blumenhagen:2017ogh} to quantum ${\cal W}$-algebras. We will see that the higher
products now involve the normal ordered product as the fundamental
one, and that they also receive $\hbar$
corrections.
In addition also the quadratic relations among the higher products
receive  quantum corrections, induced by non-trivial contractions
following from the application of Wick's theorem.
Since we are dealing with an interacting (non-free) CFT, these
contractions are given by the singular part of the operator product
expansion (OPE) and, as will be shown, imply off-diagonal terms among
the naive classical L$_\infty$ relations. Guided by quantum ${\cal W}$
algebras we are thus led to a well motivated definition of quantum
L$_\infty$ algebras that control the symmetries of a quantum theory. Similar as in the case of classical symmetries the quantum L$_\infty$ algebras we look at are restricted to a graded vector space $X = X_0 \oplus X_{-1}$ and are constructed such that they become the classical L$_\infty$ algebra of the classical symmetry in the $\hbar \rightarrow 0$ limit.

The paper is organized as follows: In section \ref{classical}  we
recall the definition of a classical  L$_\infty$ algebra and its
connection to the gauge algebra of classical gauge field theories. In
section \ref{quantum}, after identifying the possible origin of
quantum corrections,  we first define quantum L$_\infty$
algebras. Then we will compare it to loop L$_\infty$ algebras, the quantum corrected L$_\infty$ algebras arising for closed string field theory (CSFT)\cite{Zwiebach:1992ie,Markl:1997bj}. In section \ref{WExample} we will show in detail that the quantum ${\cal W}_3$ algebra is organized in terms of a quantum L$_\infty$ algebra.


\section{The L$_\infty$ gauge algebra of a classical
  symmetry} \label{classical}

In this section we review how a perturbative classical gauge algebra is actually controlled by an L$_\infty$ algebra. L$_\infty$ algebras are generalized Lie algebras where one has not only a two-product, the commutator, but more general multilinear $n$-products with $n$ inputs
\eq{
\ell_n: \qquad \quad X^{\otimes n} &\rightarrow X \\
x_1, \dots , x_n &\mapsto \ell_n(x_1, \dots , x_n) \, , 
}
defined on a graded vector space $X = \bigoplus_n X_n$, where $n$ denotes the grading.
The products are graded symmetric 
\eq{ 
\label{permuting}
\ell_n (\dots, x_1,x_2, \dots) = (-1)^{1+ {\rm deg}(x_1) {\rm deg}( x_2)} \, \ell_2 (\dots, x_2,x_1, \dots )\,,
}
with 
\eq{
      {\rm deg}\big( \, \ell_n(x_1,\ldots,x_n)\, \big)=n-2+\sum_{i=1}^n  {\rm deg}(x_i)\,.
}
 These $\ell_n$ define an L$_\infty$ algebra, if they satisfy the
 infinitely many relations
\eq{
\label{linftyrels}
{\cal J}_n(x_1,\ldots, x_n):=\sum_{i + j = n + 1 } &(-1)^{i(j-1)} \sum_\sigma \chi (\sigma;x) \; \\
 &\ell_j \big( \;
\ell_i (x_{\sigma(1)}\; , \dots , x_{\sigma(i)} )\, , x_{\sigma(i+1)} , \dots ,
x_{\sigma(n)} \, \big) = 0 \, .
}
The permutations are restricted to the ones with
\eq{ 
\label{restrictiononpermutation}
\sigma(1) < \cdots < \sigma(i) , \qquad \sigma(i+1) < \cdots < \sigma(n)\,,
}
and the sign $\chi(\sigma; x) = \pm 1$ can be read off from \eqref{permuting}. 
The first relations ${\cal J}_n$ with $n=1,2,3,\ldots$ can be schematically written as 
\eq{ 
&{\cal J}_1 = \ell_1 \ell_1 \, , \qquad {\cal J}_2 = \ell_1 \ell_2 - \ell_2 \ell_1 \,,  \qquad {\cal J}_3 =
\ell_1 \ell_3 + \ell_2 \ell_2 + \ell_3 \ell_1 \,, \\[0.1cm] &
 {\cal J}_4 = \ell_1 \ell_4 - \ell_2 \ell_3 + \ell_3 \ell_2 - \ell_4 \ell_1 \, , 
}
from which one can deduce the scheme for the higher ${\cal J}_n$. More concretely, the first L$_\infty$ relations read
\eq{   \label{ininftyrel1}
\ell_1\big( \,  \ell_1   (x) \, \big) &= 0 \\
\ell_1 \big( \, \ell_2(x_1, x_2)\,\big) &=  \ell_2\big(\,  \ell_1 (x_1) , x_2 \, \big) + (-1)^{x_1} \ell_2\big(\, x_1, \ell_1 (x_2) \, \big) \, ,
}
revealing that $\ell_1$ must be a nilpotent derivation with respect to
$\ell_2$. Denoting $(-1)^{x_i}=(-1)^{deg(x_i)}$ the full relation $ {\cal J}_3 $ reads
\begin{eqnarray}    \label{ininftyrel2}
       0\!\!\!&=&\!\!\! \phantom{} \ell_1\big(\ell_3(x_1,x_2,x_3)\, \big) +\\[0.2cm]
     &&\ell_2\big(\ell_2(x_1,x_2),x_3\, \big)+(-1)^{(x_2+x_3)x_1}
     \ell_2\big(\ell_2(x_2,x_3),x_1\, \big)+\nonumber \\[0.1cm]
   &&(-1)^{(x_1+x_2)x_3}
     \ell_2\big(\ell_2(x_3,x_1),x_2\, \big)+\nonumber \\[0.2cm]
     &&\ell_3\big(\ell_1(x_1),x_2,x_3\, \big)+(-1)^{x_1}\ell_3\big(x_1,\ell_1(x_2),x_3\, \big)
+(-1)^{x_1+x_2}\ell_3\big(x_1,x_2,\ell_1(x_3)\, \big)\, \nonumber
\end{eqnarray}   
and means that 
the Jacobi identity for the $\ell_2$ product is mildly violated by
$\ell_1$ exact expressions.
For this reason,  L$_\infty$ algebras are also
called strong homotopy Lie algebras in the mathematical literature. 

The framework of L$_\infty$ algebras is quite flexible
and it has been suggested  that every classical perturbative gauge
theory (derived from string theory),
including its dynamics, is organized by an underlying L$_\infty$
structure \cite{Hohm:2017pnh}. For sure, the pure gauge algebra of such theories satisfies
the L$_\infty$ identities.  To see this, let us assume that the field theory has a standard
type gauge structure, meaning that the variations of the fields can be
organized unambiguously into a sum of terms each of a definite power in the fields. Defining the space of gauge parameters $\varepsilon$ to be
$X_0$ and the field space $\Phi$ to be $X_{-1}$ and setting all other
graded vector spaces to be trivial, the gauge variations
can be expressed  as
\eq{\label{var}
  \delta_{\varepsilon} \Phi &=\sum_{n\ge 0}   {1\over n!}
      (-1)^{n(n-1)\over 2}\,
 \ell_{n+1}(\varepsilon, \underbrace{ \Phi, \dots, \Phi}_{n \; {\rm times}} )\, .
 }
It was shown in \cite{Hohm:2017pnh,Berends:1984rq,Fulp:2002kk,Burgers:diss}, that the closure of the symmetry
variations 
\eq{
\label{commurel4}
                      [\delta_{\varepsilon_1},\delta_{\varepsilon_2}] \Phi
                      =\delta_{- C(\varepsilon_1,\varepsilon_2, \Phi)} \Phi \, ,
}
and the Jacobi identity
 \eq{ \label{Gaugejacobiator4}
\sum_{\rm cycl} \big[ \delta_{\varepsilon_1}, [  \delta_{\varepsilon_2} ,  \delta_{\varepsilon_3} ] \big] = 0 \,  
}
are equivalent to the L$_\infty$ relations with two and three gauge parameters.
Here the  closure relation allows for a field dependent gauge parameter which can be written in terms of L$_\infty$ products as
\eq{\label{clclosure}
C(\varepsilon_1,\varepsilon_2, \Phi) & =\sum_{n\ge 0} {1\over n!}
            (-1)^{{n(n-1)\over 2}}
            \, \ell_{n+2}(\varepsilon_1,\varepsilon_2,  \underbrace{ \Phi, \dots, \Phi}_{n\;  {\rm times}})\,.
            }
Since it is precisely these relations that we will extend to the
quantum case, let us briefly exemplify the procedure
of identifying the constraints arising from the gauge closure with
L$_\infty$ relations up to cubic order in the fields. Using
\eqref{var}, the gauge commutator
 reads
\eq{ \label{cretecommu}
[ \delta_{\varepsilon_1}, \delta_{\varepsilon_2} ] \Phi = & \phantom{.} \Big\{\ell_2\big( \, \varepsilon_2, \ell_1(\varepsilon_1)\, \big)  \\[0.2cm] &
+ \ell_2\big( \,\varepsilon_2, \ell_2 (\varepsilon_1, \Phi) \, \big) - \ell_3 \big( \, \varepsilon_2, \ell_1(\varepsilon_1), \Phi \, \big) \, \\[0.2cm]
&  - \ell_3 \big( \,\varepsilon_2, \ell_2(\varepsilon_1, \Phi), \Phi \, \big) - \tfrac{1}{2} \ell_2\big( \, \varepsilon_2, \ell_3(\varepsilon_1, \Phi,\Phi) \, \big) \Big\}\\[0.2cm]
&- \Big\{\varepsilon_1 \leftrightarrow \varepsilon_2\Big\} + {\cal O}(\Phi^3) \,  ,
}
while the right hand side of the gauge closure condition can be 
expanded as
\begin{eqnarray}
\label{concreteclosure}
\delta_{-C(\varepsilon_1, \varepsilon_2, \Phi) } \Phi \!\!\!&=& \!\!\!\delta_{- \ell_2(\varepsilon_1, \varepsilon_2 )} \Phi + \delta_{- \ell_3(\varepsilon_1, \varepsilon_2, \Phi) }\Phi + \delta_{{1\over 2}\ell_4(\varepsilon_1, \varepsilon_2, \Phi, \Phi) } 
\Phi + {\cal O} (\Phi^3) \nonumber\\[0,2cm]
&& = -\ell_1 \big( \, \ell_2( \varepsilon_1, \varepsilon_2) \, \big) - \ell_2 \big( \, \ell_2(  \varepsilon_1, \varepsilon_2 ) ,  \Phi \, \big) + \tfrac{1}{2} \ell_3 \big( \, \ell_2( \varepsilon_1, \varepsilon_2) , \Phi, \Phi \, \big) \nonumber\\[0,2cm]
&&\phantom{=} \;\, - \ell_1 \big( \, \ell_3( \varepsilon_1, \varepsilon_2, \Phi) \, \big) - \ell_2 \big( \, \ell_3  (\varepsilon_1, \varepsilon_2, \Phi ) , \Phi \, \big) \\[0,2cm]
&&\phantom{=} \;\, +{1\over 2} \ell_1(\ell_4(\varepsilon_1, \varepsilon_2, \Phi , \Phi) \, \big) +{\cal O}(\Phi^3)\,.\nonumber
\end{eqnarray}
Comparing \eqref{concreteclosure} with \eqref{cretecommu} we see that
demanding closure yields conditions on the $\ell_n$ products. For
instance, at zeroth  order in $\Phi$ one obtains the condition
\eq{ \label{J2}
\ell_1 \big( \,\ell_2(\varepsilon_1, \varepsilon_2)\, \big) = \ell_2\big( \,\varepsilon_1, \ell_1 (\varepsilon_2) \, \big) - \ell_2\big( \,\varepsilon_2, \ell_1 (\varepsilon_1) \, \big) \, .
}
Upon interchanging the arguments this is exactly the L$_\infty$
relation ${\cal J}_2(\varepsilon_1, \varepsilon_2) = 0$ in
\eqref{ininftyrel1}. At first order in $\Phi$ one gets  
\eq{ \label{J3}
0 =\,& \ell_2\big( \,\varepsilon_2, \ell_2 (\varepsilon_1, \Phi) \, \big)  + \ell_2 \big( \, \ell_2(  \varepsilon_1, \varepsilon_2 ) ,  \Phi \, \big) -  \ell_2\big( \,\varepsilon_1, \ell_2 (\varepsilon_2, \Phi) \, \big) \\
& - \ell_3 \big( \, \varepsilon_2, \ell_1(\varepsilon_1), \Phi \, \big)  + \ell_3 \big( \, \varepsilon_1, \ell_1(\varepsilon_2), \Phi \, \big) \\&+ \ell_1 \big( \, \ell_3( \varepsilon_1, \varepsilon_2, \Phi) \, \big) \, .
}
This is the L$_\infty$ relation ${\cal J}_3(\varepsilon_1,
\varepsilon_2, \Phi) = 0$ in which the term $\ell_3\big(
\,\varepsilon_1, \epsilon_2 , \ell_1 (\Phi) \, \big)$ is missing, as
we have set $X_{-2}=0$.
 This result is just a consequence of the general  two relations between the classical gauge algebra and  the L$_\infty$ algebra:
\eq{
\text{gauge closure} \qquad \Leftrightarrow \qquad 0 =& \, {\cal J}_n(\varepsilon_1, \varepsilon_2, \underbrace {\Phi, \dots, \Phi}_{n-2 \, {\rm times}  } ) \, ,
}

\vspace{-0.5cm}
\eq{
\text{gauge Jacobi identity} \qquad \Leftrightarrow \qquad  0 =& \,  {\cal J}_n(\varepsilon_1, \varepsilon_2, \varepsilon_3,  \underbrace {\Phi, \dots, \Phi}_{n-3 \, {\rm times}  })\, . 
}
As one can check, these are actually the only non-trivial L$_\infty$
relations in case that the graded vector space is given by $X=X_0\oplus
X_{-1}$. 
This can be generalized by adding a vector space $X_{-2}$ containing the equations
of motion, thus allowing the freedom that  gauge closure
only holds on-shell \cite{Hohm:2017pnh}.

\section{Quantum L$_\infty$ gauge algebras} \label{quantum}

In the last section we recalled  how the L$_\infty$ relations
guarantee the consistency of a classical gauge algebra. Recently it
was shown that also global classical ${\cal W}$ algebras arising in   two-dimensional
conformal field theory yield  non-trivial examples of L$_\infty$ 
algebras. Driven by the aim to extract  physically well motivated
aspects of  a quantum extension of  L$_\infty$ algebras,
we analyze  whether a
generalized version of this correspondence holds for quantum
${\cal W}$ algebras. 
On the way, we encounter a couple of new structures  that can be
traced back to the non-associativity of the normal ordered products
appearing in the quantum ${\cal W}$ algebra.
Resolving these issues  guides us to a proposal of a quantum L$_\infty$ 
gauge algebra that we will present in the section.

Concretely, in section \ref{quentumlinfty},  by demanding consistency of the quantized symmetry
algebra, we  outline how the usual
notion of an L$_\infty$ algebra has to be adjusted for a quantum
L$_\infty$ algebra. We find that beyond  the higher products also the
L$_\infty$ relations receive quantum corrections, whose origin lies in
the necessity to perform  Wick contractions between quantum fields.

In \ref{CSFT} we review the L$_\infty$ algebra of closed string field theory and the quantum corrections appearing there. As it turns out, the quantum corrections due to Wick contractions do not appear there.

\subsection{The quantum L$_\infty$ algebra of a quantum
  symmetry} \label{quentumlinfty}

Going from a classical field theory to a quantum field theory, the
fields become operator valued. We want to consider quantum
symmetries which in the classical limit $\hbar \rightarrow 0$ become  a
classical symmetry of the kind described in the last section. In
particular we are still working only on the graded vector space
$X=X_0\oplus X_{-1}$, where the symmetry parameters are contained in
$X_0$ and the field operators in $X_{-1}$. In the case of ${\cal
  W}$-algebras, the infinitely many symmetry parameters\footnote{Note that the holomorphic function
  $\epsilon(z)$ 
does not parametrize a gauge variation, as the latter would depend on
$z$ and $\ov{z}$.}
are compactly
encoded in  $\epsilon(z)=\sum_{n\in \mathbb Z} z^{n+\Delta-1} \epsilon_n$ and
the infinitely many symmetry generators in $W(z)=\sum_{n\in \mathbb Z}
z^{-n-\Delta} W_n$. Here $\Delta$ denotes the conformal dimension of
the chiral field $W(z)$.

In the classical case it
was crucial that the variation of the field could be organized in terms
of definite powers in the fields to define the corresponding L$_\infty$
products. In order to adapt the notion of field powers, we have to
specify an operator product in the quantum case. 

Inspired by the analysis of ${\cal W}$ algebras, to be discussed in detail in section
\ref{WExample},  we define the operator product to be
the symmetrized normal ordered 
product
\eq{ \label{starproduct}
A \star B = {1 \over 2} \Big( \, N(AB) + N(BA) \, \Big) \, .
}
This is a convenient choice, as by taking the classical limit $\hbar
\rightarrow 0$, it becomes the usual point-wise multiplication of
fields. Let us already point out one subtlety relative to the
classical case, that will be one source of quantum corrections.
As can be seen from the notion of the normal ordering in 2d CFT,
the $\star$ product above while commutative fails to be 
associative.
There\footnote{This can be shown using the general formula 6.227
  in \cite{di1997conformal}.}, 
the non-associativity of the normal ordered product  is given by
\eq{ \label{nonassociativity}
 (\varepsilon   A) \star  B -   \varepsilon  ( A \star B) =  \varepsilon   (\contraction{}{A}{}{B}AB )\, ,
}
where $\varepsilon$ is just a c-number symmetry variation and $A,B$ are
operator valued fields.
Moreover, the last term denotes extra terms arising from the contraction between the two
  operators defined as
\eq{
     \lim_{y\to x}  \Big( A(x)\, B(y) - (\contraction{}{A}{}{B}AB )(x,y)\Big)=N(A\,B)(x)
}
which in a CFT is nothing else than the singular part of the operator
product expansion.
Having defined the product between operators, 
we assume that variations of the field can be schematically written in the form
\eq{ \label{formofexpansion}
\delta_\varepsilon^{\rm qu} \, \Phi    \sim \sum_n  \varepsilon \;  \underbrace{\Phi \star \dots  \star \Phi }_{n \; {\rm times}} \, ,
}
where for simplicity we considered  bosonic  fields and  symmetry parameters. Following the lines of the classical discussion we define graded symmetric multilinear quantum n-products
\eq{
L_{n+1}:\, X^{\otimes n}\rightarrow X
}
and rewrite the variation in the form
\eq{
\label{expan1quant}
        \delta_{\varepsilon}^{\rm qu} \,  \Phi =\sum_{n\ge 0}   {1\over n!}
      (-1)^{n(n-1)\over 2}\,
 L_{n+1}(\varepsilon, \underbrace{ \Phi, \dots, \Phi}_{n \; {\rm times}} )\, . 
}
The quantum $L_n$ products still carry the intrinsic grading ${\rm
  deg}\,L_n=n-2$. Since the star-product is symmetric, the
$L$-products are automatically symmetric when interchanging two
fields. Since in the limit $\hbar \rightarrow 0$,  the star product becomes the normal field product, the quantum $L_n$-products will
become the classical $\ell_n$-products with the right degree and
symmetry properties.

Following the classical analysis, the question now is which constraints arise from demanding the closure of the quantum symmetry algebra
\eq{
\label{commurelquant}
                      [\delta_{\varepsilon_1}^{\rm qu} \, ,\delta_{\varepsilon_2}^{\rm qu} \, ] \Phi
                      =\delta_{- C \,(\varepsilon_1,\varepsilon_2, \Phi)}^{\rm qu} \, \Phi \,
}
and the  Jacobi identity 
\eq{ \label{Gaugejacobiator}
\sum_{\rm cycl} \big[ \delta_{\varepsilon_1}^{\rm qu} \,, [  \delta_{\varepsilon_2}^{\rm qu} \, ,  \delta_{\varepsilon_3}^{\rm qu} \, ] \big] = 0 \,.  
}
Here, the field dependent closure parameter $ C(\varepsilon_1,\varepsilon_2, \Phi)$ should still be expressed 
in terms of the symmetrized normal ordered product  
\eq{
 C\,(\varepsilon_1,\varepsilon_2, \Phi)\sim \sum_n \varepsilon_1  \varepsilon_2 \cdot \Phi \star \dots \star  \Phi
\,,} 
allowing  to read off the $L_n$ products with two symmetry parameters
\eq{
\label{expan2}
           C\, (\varepsilon_1,\varepsilon_2, \Phi) =\sum_{n\ge 0} {1\over n!}
            (-1)^{{n(n-1)\over 2}}
            \;L_{n+2}(\varepsilon_1,\varepsilon_2,  \underbrace{ \Phi, \dots, \Phi}_{n\;  {\rm times}})\,.
}
To identify potential sources of quantum corrections in the L$_\infty$  relations,
we write out the first few terms of both sides of the closure
condition \eqref{commurelquant}.
Up to second order in the fields, the left hand side can be expanded as
\begin{eqnarray}
\label{quantumclosure}
[ \delta_{\varepsilon_1}^{\rm qu}\,,\, \delta_{\varepsilon_2}^{\rm qu}\,
] \Phi \!\!\!& = &\!\!\! \phantom{} \Big\{L_2\big( \, \varepsilon_2,
L_1(\varepsilon_1)\, \big)  \nonumber\\[0.2cm] 
&&\!\!\!\!\!\!\! + L_2\big( \,\varepsilon_2, L_2 (\varepsilon_1, \Phi) \, \big) - L_3 \big( \, \varepsilon_2, L_1(\varepsilon_1), \Phi \, \big) \, \\[0.2cm]
&& \!\!\!\!\! \!\! -L_3 \big( \,\varepsilon_2, L_2(\varepsilon_1, \Phi), \Phi \, \big)- \tfrac{1}{2} L_2\big( \, \varepsilon_2, L_3(\varepsilon_1, \Phi,\Phi) \, \big) \Big\} \nonumber \\[0.2cm]
&&\!\!\!\!\!\!\! - \Big\{\varepsilon_1 \leftrightarrow \varepsilon_2\Big\}\,  ,\nonumber
\end{eqnarray}
while the right side is
\begin{eqnarray} 
\label{quantumexample}
\delta_{-C^{\rm qu}\,(\varepsilon_1, \varepsilon_2, \Phi) } \Phi \!\!\!&=&\!\!\! \delta_{- L_2(\varepsilon_1, \varepsilon_2 )} \Phi + \delta_{- L_3(\varepsilon_1, \varepsilon_2, \Phi) } \Phi\\[0,2cm]
& =&\!\!\! -L_1 \big( \, L_2( \varepsilon_1, \varepsilon_2) \, \big) - L_2 \big( \, L_2(  \varepsilon_1, \varepsilon_2 ) ,  \Phi \, \big) + \tfrac{1}{2} L_3 \big( \, L_2( \varepsilon_1, \varepsilon_2) , \Phi, \Phi \, \big) \nonumber\\[0,2cm]
&&\!\!\! - L_1 \big( \, L_3( \varepsilon_1, \varepsilon_2, \Phi) \, \big) - L_2 \big( \, L_3  (\varepsilon_1, \varepsilon_2, \Phi ) , \Phi \, \big)\, . \nonumber
\end{eqnarray}
To read off the quantum L$_\infty$ relations, we now sort
\eqref{quantumclosure} and \eqref{quantumexample} according to the
power in $\Phi$ . 
Since now the power of $\Phi$ is with respect to  the symmetrized normal
ordered product, this is a bit more subtle than in the classical case.
One first has to bring all terms into the schematic form
$(\varepsilon_1  \varepsilon_2) \cdot (\Phi \star \dots \star \Phi)$
that also appeared in the definitions of the $L$-products
\eqref{expan1quant} and \eqref{expan2}.
While some terms are already of this form, for others  a rebracketing
is necessary. 

Consider for instance the fourth term in \eqref{quantumclosure} that,
upon using \eqref{formofexpansion}, 
can be schematically written as
\eq{
L_3 \big( \,\varepsilon_2, L_2(\varepsilon_1, \Phi), \Phi \, \big) \sim \varepsilon_2  \big( \,  ( \varepsilon_1  \Phi) \star  \Phi \, \big)\,.
}
Using the non-associativity of the $\star$-product \eqref{nonassociativity}, this becomes
\eq{ \label{secondterm}
L_3 \big( \,\varepsilon_2, L_2(\varepsilon_1, \Phi), \Phi \, \big) = \varepsilon_1 \varepsilon_2 \cdot  (\Phi \star \Phi) + \varepsilon_1 \varepsilon_2 \cdot  (\contraction{}{\Phi}{}{\Phi}\Phi  \Phi)\, . 
}
Let us  assume for simplicity a free theory such that
$\contraction{}{\Phi}{}{\Phi}\Phi  \Phi $ is proportional $\hbar
\mathbf{1}$. Then the last term in \eqref{secondterm} is proportional
to $\epsilon_1$ and $\epsilon_2$ and therefore a quantum correction to
the L$_\infty$ relation at zeroth order in $\Phi$. Treating the last
term in \eqref{quantumexample} in an analogous way, we find the quantum corrected L$_\infty$ relation at zeroth order in $\Phi$
\eq{
0 &=  L_2\big( \, L_1(\varepsilon_1) , \varepsilon_2\, \big) + L_2\big( \,\varepsilon_1, L_1(\varepsilon_2 ) \, \big) +  L_1\big( \,L_2(\varepsilon_1, \varepsilon_2) \, \big) \\
 &- L_3 \big( \,\varepsilon_2, L_2(\varepsilon_1, \contraction{}{\Phi}{),}{\Phi} \Phi), \Phi \, \big) + L_3 \big( \,\varepsilon_1, L_2(\varepsilon_2, \contraction{}{\Phi}{),}{\Phi} \Phi), \Phi \, \big) + L_2 \big( \, L_3  (\varepsilon_1, \varepsilon_2, \contraction{}{\Phi}{),}{\Phi}{} \Phi ) , \Phi \, \big) \, . 
}
Similarly also all other L$_\infty$ relations get corrected  by contractions of higher L$_\infty$ relations.

Let us summarize: Guided by quantum algebras in 2d CFT, we identified
two sources of quantum corrections to L$_\infty$ algebra. First,
relative  to the classical products, the
higher quantum L$_\infty$ products can receive corrections of
higher order in $\hbar$ . The second kind of quantum corrections
arises from contractions between quantum fields that appear 
when sorting the relations in powers of the field. 
These contractions change the power of the fields  so that the
classically separated L$_\infty$ relations receive quantum suppressed
off-diagonal corrections.

We want to stress that the contractions differ severely from theory to
theory. While in free theories the contraction is proportional to the
identity operator, in interacting theories (like generic CFTs) the contraction of two
fields is usually field dependent again. We can therefore not provide
a general closed formula for   which contraction of which L$_\infty$ relation
contributes to which other L$_\infty$ relation. 

Guided by these
observations we suggest to define quantum L$_\infty$ algebras that
govern (global) quantum symmetries as 
follows:
One has a graded vector space 
$X= X_0 \oplus X_{-1}$, where $X_n$ is said to have
degree $n$. In addition there are multi-linear quantum products
$L_n(x_1,\ldots,x_n)$ that have degree ${\rm deg}(L_n)=n-2$
so that
\eq{
      {\rm deg}\big( \, L_n(x_1,\ldots,x_n)\, \big)=n-2+\sum_{i=1}^n  {\rm deg}(x_i)\,.
}
Each product can in principle receive quantum corrections at any  power in $\hbar$.
The products are graded commutative, i.e.
\eq{ 
\label{permuting2}
L_n (\dots, x_1,x_2, \dots) = (-1)^{1+ {\rm deg}(x_1) {\rm deg}( x_2)} L_n (\dots, x_2,x_1, \dots )\,.
}
Like in the classical case, one  defines
\eq{
\label{linftyrels2}
{\cal J}^{\rm qu}_n(x_1,\ldots, x_n):=\sum_{i + j = n + 1 } &(-1)^{i(j-1)} \sum_\sigma 
            \chi(\sigma;x) \; \\
 &L_j \big( \, \;
L_i (x_{\sigma(1)}\; , \dots , x_{\sigma(i)} )\, , x_{\sigma(i+1)} , \dots ,
x_{\sigma(n)} \, \big)  \,.
}\newpage The $L_n$ products define a quantum L$_\infty$ algebra if they satisfy for each $m = 2,3$ and  $n\in \mathbb{Z}^{+}_0$ 
\eq{
\label{quantumrelas}
          {\cal J}^{\rm qu}_{m+n}(\epsilon_1, \dots , \epsilon_m, x_1,\ldots, x_n)+\sum_{\substack{ (y_1,\ldots, y_k) \\ \rightarrow (x_1, \dots, x_n) }}  \hbar^\xi \; {\cal
            J}^{\rm qu}_{m+k}(\epsilon_1, \dots,  \epsilon_m, \underbrace{y_1,\ldots, y_k}_{\to (x_1,\ldots,x_n)})=0
}
Since this is the main formula of the paper we want to explain the
formula in more detail. $\epsilon_i \in X_0$ is a symmetry parameter
and $x_i \in X_{-1}$ is a field. While the first term is the known one
from the classical L$_\infty$ relations, the second term contains the crucial new feature of quantum L$_\infty$ algebras, namely the corrections due to contractions of other L$_\infty$ relations. To cover all such corrections we sum over all L$_\infty$ relations whose field input $(y_1, \dots , y_k)$ can contract into $(x_1, \dots , x_n)$. The $\xi\ge 1$ counts the number of contractions employed to convert 
the dependence on $(y_1,\ldots, y_m)$ into a dependence on $(x_1,\ldots, x_n)$. The underbrace signals that only the terms that arise from the particular contraction are to be taken here. To avoid permutation factors we let the sum run only over $(y_1, \dots , y_k)$ that are not equal under permutation. Furthermore notice that the order of the $(y_1, \dots , y_k)$ does not play a role since the ${\cal J}^{\rm qu}$ share the permutation property of \eqref{permuting2}
\footnote{Here an obstacle becomes apparent if one tries to generalize the above definition beyond the given case where contractions appear only between elements of $X_{-1}$. When contractions appear not only between elements with even parity the order of the $y_1, \dots , y_k$ does indeed matter. Lacking an example to follow we cannot give a precise ordering prescription to fix this issue here. }.

Let us provide a more general and mathematically precise definition for the quantum L$_\infty$ algebra. Since the quantum corrections mix the different L$_\infty$ relations, we can also define quantum L$_\infty$ algebras very compactly by demanding that for $m\in \left\lbrace 2,3\right\rbrace$ and $\epsilon_i \in X_0$ the sum of all  L$_\infty$ relations vanish
\eq{ \label{quantumLprecise}
                       \sum_{n=1}^\infty \; \sum_{ (x_1, \dots x_n) \in X_{-1}^n  } {\cal J}^{\rm qu}_{m+n} (\epsilon_1, \dots, \epsilon_m, x_1, \dots x_n)=0 \, , 
} 
where as before the second sum runs only over distinct $(x_1, \dots, x_n)$. In case the L products do not change the power of the input, the terms in \eqref{quantumLprecise} separate into the classical L$_\infty$ relations \eqref{linftyrels}. On the other hand, using normal ordered products in the L products, \eqref{quantumLprecise} reduces to the former definition \eqref{quantumrelas}. Nevertheless we want to stress that in general \eqref{quantumLprecise} does not need any physical input in form of a contraction. From the mathematical viewpoint the definition \eqref{quantumLprecise} might therefore be more appealing. We nevertheless prefer \eqref{quantumrelas} that also makes it manifest
that in the $\hbar\to 0$ limit one encounters the classical L$_\infty$
relations and that their off-diagonal quantum corrections arise from the
contraction of quantum fields.

In section \ref{WExample} we show in much detail 
how quantum ${\cal W}$ algebras fit precisely into this definition of
quantum L$_\infty$ algebras. Especially in section
\ref{checkingrelations} we will demonstrate that the quantum relations
\eqref{quantumrelas} can be given a precise meaning for the quantum ${\cal W}_3$ algebra.

\subsection{Comparison to the L$_\infty$ algebra of CSFT} \label{CSFT}

We will now compare our definition for a quantum L$_\infty$ algebra
with the L$_\infty$ algebra of closed string field theory (CSFT)
\cite{Zwiebach:1992ie,Markl:1997bj}. To distinguish these two
different L$_\infty$ definitions, we will follow Markl \cite{Markl:1997bj} and call the
L$_\infty$ algebra of CSFT a loop L$_\infty$ algebra, while the
definition from last section will be called quantum L$_\infty$ algebra.

In a loop L$_\infty$ algebras one usually expands the quantum products according to their loop level, thus their power of $\hbar$
\eq{ \label{hbarexp}
L_n (x_1, \dots , x_n) = \sum_g  L^g_n (x_1, \dots x_n) \, , 
}
where $L_n^g$ is proportional to $\hbar ^g$. 
Then, the $L_n^g$ products define a loop L$_\infty$ algebra, if for any level $g$ the following relation holds (we use the notation of \cite{Markl:1997bj})
\eq{  \label{quantumfundamental}
0 =& \sum_{g_1 + g_2 = g} \,  \sum_{i + j = n + 1 } (-1)^{i(j-1)} \sum_\sigma 
            \chi(\sigma;x) \; \\
 & \qquad \qquad \qquad \times L_j^{g_1} \big( \;
L_i^{g_2} (x_{\sigma(1)}\; , \dots , x_{\sigma(i)} )\, , x_{\sigma(i+1)} , \dots ,
x_{\sigma(n)} \, \big)  \, \\
& + {1 \over 2} \sum_s (-1)^{ {\rm deg}(h_s) + n - g} \,L_{n+2}^{g-1} (h_s, h^s, x_1, \dots , x_n) \, . 
}
The sum over $s$ in the last term runs over a basis of fields labeled
by $s$. The field with an upper index, $h^s$, is the conjugate field
to $h_s$ with respect to a scalar product $\langle h^s, h_t \rangle =
\delta^s_t$. The $\sum_s L_n^{g-1}(h_s, h^s, \dots)$ can be
interpreted as an identity operator. When contracting $h_s, h^s$ to
eliminate this identity operator, we obtain an additional $\hbar$
factor such that, together with the $\hbar^{g-1}$ from the
$L_n^{g-1}$, the last term is proportional to $\hbar^g$ as well.

 Let us compare the defining relations of {(global) quantum and
   (gauge) loop} L$_\infty$ algebras:
The first part of the loop L$_\infty$ relation \eqref{quantumfundamental} appears in quantum L$_\infty$ algebras as the order $\hbar^g$ term, when inserting the expansion \eqref{hbarexp} into the first term of \eqref{quantumrelas}.
The second term of \eqref{quantumfundamental} does not appear in the
quantum L$_\infty$ relations in \eqref{quantumrelas}. The reason for
this is, that the quantum L$_\infty$ was derived in a setting where
the total vector space contained only degree 0 objects, the symmetry
parameters, and degree -1 objects, the fields. Therefore $X = X_0
\oplus X_{-1}$ and all objects with a degree other than 0 and -1 were
set to  zero. Demanding all terms in the defining relation of loop L$_\infty$ algebras \eqref{quantumfundamental} to have the same degree, we find
\eq{
{\rm deg}(h_s) + {\rm deg}(h^s) = -3 \, . 
}
Since $h_s$ is a field, its degree is ${\rm deg}(h_s) = -1$ and the
degree of $h^s$ is bound to be ${\rm deg}(h^s) = -2$. Therefore,
$h^s$ is  trivial 
and the second term in \eqref{quantumfundamental} could  not appear in
the derivation of the quantum L$_\infty$ based entirely on  quantum
gauge variations.

Remarkably, the second term in the quantum L$_\infty$ relation
\eqref{quantumrelas} has no counterpart in the loop L$_\infty$
algebras.  Therefore the L$_\infty$ relations of the CSFT L$_\infty$
algebra do not receive corrections 
from contraction terms. 
The question arises if there exist  a connection between the two
definitions. From the current status, the answer is not completely clear
to us and more work or insight is required to fully clarify it.
We can only say that  the structure of  (gauge) loop L$_\infty$ arose as a
consequence of the quantum master equation of the BV-formalism for the
CSFT quantum action. On the contrary,  our (global) quantum L$_\infty$
definition is based on the analysis of 
bootstrapped and therefore exactly solvable global quantum ${\cal W}$
algebras in 2d CFT.

\section{The quantum ${\cal W}_3-{\rm L}_{\infty}$ algebra} 
\label{WExample}

In the recent paper \cite{Blumenhagen:2017ogh} it was shown that
(classical) ${\cal W}$ algebras are highly non-trivial (classical)
L$_\infty$ algebras with field dependent symmetry  parameters. 
In this section we will show that the quantum ${\cal W}_3$-algebra
fits into the framework of the quantum L$_\infty$ algebra of section
\ref{quentumlinfty} (and was in fact motivating it).  We expect that more general
quantum ${\cal W}$-algebras 
will even provide more intricate examples of  quantum L$_\infty$ algebras.

\subsection{ ${\cal W}$ algebras}

In two-dimensional conformal field theories the energy momentum tensor
$T(z)$ is a quasi primary field that has conformal dimension two,
generates the conformal transformations and obeys the Virasoro
algebra. A ${\cal W}$ algebra is an extension of the Virasoro algebra
by chiral primary fields of conformal dimension usually larger than two.
The prototype example is Zamolodchikov's ${\cal W}_3$ algebra \cite{Zamolodchikov:1985wn},
generated by two fields $\{T(z),W(z)\}$ of conformal dimensions two
and three. 
The (quantum) OPEs among these fields are known to be\footnote{Up to
  some structure constants, the
   form of the OPE between quasi-primary fields is generally known
   \cite{Blumenhagen:1990jv} (for a pedestrian derivation see also
   \cite{Blumenhagen:2009zz}), 
    as has been exploited for the classical ${\cal W}-{\rm  L}_\infty$ algebra 
   relation in \cite{Blumenhagen:2017ogh}.}
\eq{ \label{OPES}
       {1\over \hbar}\,  T(z)\circ T(w)&={c/2\over (z-w)^4} +2\left( {T(w)\over
             (z-w)^2}+{1\over 2} {\partial T(w)\over
             (z-w)}\right)\,,\\[0.3cm]
    {1\over \hbar}\,    T(z)\circ W(w)&=3\left( {W(w)\over
             (z-w)^2}+{1\over 3} {\partial W(w)\over
             (z-w)}\right)\,,\\[0.3cm]
   {1\over \hbar}\,  W(z)\circ W(w)&={c/3\over (z-w)^6} \\[0.2cm]
             +\alpha&\left( {T(w)\over
             (z-w)^4}+{1\over 2} {\partial T(w)\over
             (z-w)^3}+{3\over 20} {\partial^2 T(w)\over
             (z-w)^2}+{1\over 30} {\partial^3 T(w)\over
             (z-w)}\right)\\[0.2cm]
             +\beta &\left( {\Lambda^{\rm qu}(w)\over
             (z-w)^2}+{1\over 2} {\partial \Lambda^{\rm qu}(w)\over
             (z-w)}\right)\,.
}
Here the field $\Lambda^{\rm qu}$ denotes the normal ordered product
\eq{
\label{quant1}
             \Lambda^{\rm qu}=N(TT)-\hbar {3 \over 10} \partial^2 T
}
where we have indicated the quantum correction linear in $T$. The
corresponding algebra for the modes satisfies the Jacobi-identity for
\eq{
\label{quant2}
                \alpha=2\,,\qquad \beta={32\over 5c+22\hbar}\,.
}
Following \cite{Bowcock:1991zk}, in these formulas we have introduced
$\hbar$ so that the classical limit and its quantum corrections are  clearly 
 visible. In the $\hbar\to 0$ limit, the commutator (singular part of the
 OPE) becomes the Poisson bracket
\eq{
   &\{\cdot,\cdot\}_{\rm PB}= \lim_{\hbar\to 0}\,  {1\over i\hbar} \,[\cdot,\cdot]\,.
}

There exist 
three sources of quantum corrections. Two of them are manifest in 
the $\hbar$ corrections in \eqref{quant1} and \eqref{quant2} \footnote{Notice that when expanding the fraction $\beta$ we get an infinite series with terms at any order in $\hbar$. Separating the different powers of $\hbar^g$ in different $L_n^g$ products, as usually done in loop L$_\infty$ algebras, see \eqref{hbarexp}, is therefore not illuminating in this example.}
 and the third is the appearance of the normal ordered product $N(TT)$ instead of the
usual point-wise multiplication $(TT)$ in the classical case.

The normal ordered product between two chiral fields is defined as
\eq{ \label{NOproduct}
               N(\phi\,\chi)(w)= {1\over 2\pi i}\oint_{\gamma(w)} dz\,
               {\phi(z)\circ \chi(w) \over (z-w)} \,, 
}
where $\gamma(w)$ is a path encircling $w$ counterclockwise once. The normal ordered product is therefore the first regular term in the OPE between the two
fields. Note that this product is neither commutative nor associative. Since for the correspondence to an L$_\infty$ algebra one needs
graded symmetric products, we use the symmetrized normal ordered
product $\star$ from \eqref{starproduct} that is still
non-associative.
To demonstrate this, let us explicitly compute the left hand side of
\eqref{nonassociativity} for $A=B=T$
\eq{
(\varepsilon T)\star T - \varepsilon (T\star T)&=
{1\over 4\pi i}\oint dz\,{\epsilon(z)\,
\contraction{}{T}{(z)\circ}{T}
 T(z)\circ T(w)
 \over (z-w)}\\[0.2cm]
&= {c\hbar\over 96}\partial^4 \epsilon+{\hbar\over 2}\partial^2 \epsilon\, T
+{\hbar\over 2}\partial \epsilon \,\partial T\,,
}
where both sides depend on $w$.
Note that these corrections arise from the  contraction of operators
below the integral and that they
are $\hbar$-suppressed relative to the leading order normal ordered products.

The extended symmetry algebra acts with
\eq{   
\label{hannover96}
 \delta_{\varepsilon_i}  W_j(w) ={1\over 2\pi i} \oint_{\gamma(w)} dz
  \, \varepsilon_i(z) \, {1\over \hbar} W_i(z)\circ  W_j(w)\, ,
} 
where $i, j = \{ T, W \}$. Instead of writing $\varepsilon_T$ and $\varepsilon_W$ from now on we will  write $\varepsilon$ for $\varepsilon_T$ and $\eta$ for $\varepsilon_W$

\subsection{$L_n$ products with one symmetry parameter}

Let us now follow the steps outlined in the sections \ref{classical}
and \ref{quentumlinfty} to construct the quantum L$_\infty$ algebra
corresponding to
the quantum ${\cal W}_3$ algebra. The fields $\{T, W\}$ have degree
$-1$, and the symmetry parameters $\{\varepsilon, \eta \}$ have degree
zero. Therefore the total vector space is $X= X_0 \oplus X_{-1}$ and each
$X_n = X_n^T \oplus X_n^W$ splits into a $T$ and a $W$ part. 
As in \cite{Blumenhagen:2017ogh}, we will
use boldface to highlight vectors in this two-dimensional space,
for instance $\textbf W = (T, W)$ will denote either of the fields. Furthermore
we equip all $L_n$ products with an upper index from the set  $\left\lbrace T, W, \epsilon, \eta\right\rbrace$
that denotes in which of the four subspaces of $X$ the image of the
higher product $L_n$ is located.

Inserting \eqref{OPES} in \eqref{hannover96},  for the  quantum corrected  infinitesimal variations one obtains
\begin{align} \label{var1}
 \delta_{\varepsilon} T&=\underbrace{{c\over 12} \, \partial^3 \varepsilon}_{L^{T}_1(\varepsilon)}
        + \underbrace{(2\, \partial \varepsilon\, T +  \varepsilon \, \partial T)}_{L^{T}_2(\varepsilon,T)} \,,\nonumber \\[0.2cm]
       \delta_{\varepsilon} W&=
        \underbrace{ (3\, \partial \varepsilon\, W +  \varepsilon
          \, \partial W)}_{L^{W}_2(\varepsilon,W)} \,,\nonumber \\[0.2cm]
        \delta_{\eta} T&=
        \underbrace{ (3\,\partial \eta\, W +  2\, \eta \, \partial W)}_{L_2^{T}(\eta,W)}\, 
        \intertext{and} \nonumber 
  \delta_{\eta} W&=\underbrace{{c\over 360} \, \partial^5
          \eta}_{L^{W}_1 (\eta)}
        + \underbrace{\alpha\Big({1\over 6} \, \partial^3 \eta\, T +
          {1\over 4} \, \partial^2
          \eta\, \partial T+ {3\over 20}\,  \partial
          \eta\, \partial^2 T+{1\over 30}\, 
          \eta\, \partial^3 T\Big)}_{L^{W}_2(\eta,T)}
        \\[0.2cm]
              &\phantom{aaaaaaaaa}-\underbrace{{3\hbar\beta\over 10}\Big(\partial
         \eta\, \partial^2 T + {1\over 2}\,  
          \eta\, \partial^3 T\Big)}_{L^{W}_{2}(\eta,T)}
        \nonumber \\[0.3cm]   
 &\phantom{=}+\underbrace{\beta \Big(\partial\eta\, (T\star T) +{1\over 2}
        \eta\, \partial (T\star T)\Big)}_{-{1\over
          2}L^{W}_3(\eta,T,T)}\,. \nonumber
\label{var2}
\end{align}
Notice that we have already written all terms in the form \eqref{formofexpansion} such that we can directly read off the $L_n$ products. 
Compared to the classical higher products, the only change is in
$\delta_{\eta} W$,
where $L^{W}_2(\eta,T)$ receives an explicit $\hbar$-correction and 
$\ell^{W}_3(\eta,T,T)$ involves the  quantum product $T\star T$.

\subsection{$L_n$ products with two symmetry  parameters}

Recall that the $L_n$ products with two symmetry parameters appear in the  closure condition \eqref{commurelquant}
\eq{ \label{closureagain}
                      [\delta_{ \mathbf{ \varepsilon}_i}^{\rm qu} \, ,\delta_{\varepsilon_j}^{\rm qu} \, ] W_k
                      =\delta_{- C(\varepsilon_i,\varepsilon_j, \mathbf W)}^{\rm qu} \, W_k \,,
}
upon expanding \eqref{expan2}
\eq{
         \mathbf  C (\varepsilon_i,\varepsilon_j, \mathbf W) =\sum_{n\ge 0} {1\over n!}
            (-1)^{{n(n-1)\over 2}}
            \;L_{n+2}(\varepsilon_i,\varepsilon_j,  \underbrace{ \mathbf W , \dots, \mathbf W}_{n\;  {\rm times}})\,.
}
To obtain the $\mathbf C (\varepsilon_i,\varepsilon_j, \mathbf W)$ we
insert \eqref{hannover96} into the symmetry  closure condition and use the generalized Wick theorem for chiral vertex operator algebras \cite{Kac:1996wd}
\begin{eqnarray}
&&\oint{dy\over 2 \pi i}\, (y-w)^n A(y)\circ\Bigg(\oint {dz\over 2 \pi
  i} \, (z-w)^m B(z)\circ C(w)\Bigg)\\ 
&&\phantom{======}-\oint{dy\over 2 \pi i}\, (y-w)^m B(y)\circ\Bigg(\oint {dz\over 2 \pi i} \, (z-w)^n A(z)\circ C(w)\Bigg)\nonumber\\[0.2cm]
&&=\sum_{j=0}^{n} \binom{n}{j}\;\oint{dz\over 2 \pi i}\Bigg( \oint{dy\over 2 \pi i} (y-z)^j\, A(y)\circ B(z)\Bigg)\circ C(w)\,(z-w)^{(m+n-j)}\nonumber
\end{eqnarray}
in the special case $m, n = 0$. In this way, for instance we can derive
\eq{
\big[\delta_{\varepsilon_1},\delta_{\varepsilon_2}\big]\,T(z)&=\left(\frac{1}{2 \pi i}\right)^2 \oint dy \,{1\over \hbar} \Bigg(\oint dw \, \varepsilon_1(w)\varepsilon_2(y)\,{1\over \hbar}T(w)\circ T(y)\Bigg)\circ\,T(z)\\
&=\frac{1}{2 \pi i}\oint dy\, \big( \,\partial \varepsilon_1(y)\varepsilon_2(y)-\varepsilon_1(y)\partial \varepsilon_2(y)\, \big)\,{1\over \hbar} T(y)\circ T(z) \, ,
} 
so that  the C-product  can be read off as  
\eq{\label{TT,T}
\mathbf{C}(\varepsilon_1,\varepsilon_2,\mathbf{W})=\varepsilon_1\, \partial \varepsilon_2-\partial \varepsilon_1\, \varepsilon_2:=L_2^{\varepsilon}(\varepsilon_1,\varepsilon_2) \, .
}
Similarly we find
\eq{\label{(TW,T)}
\mathbf{C}(\varepsilon, \eta, \mathbf{W}) &=\varepsilon\, \partial \eta-2\,\partial \varepsilon\, \eta :=L_2^{\eta}(\varepsilon,\eta)\, , \\[0.1cm]
\mathbf{C}(\eta_1,\eta_2,\mathbf{W})&=L_2^{\varepsilon}(\eta_1,\eta_2)+L_3^{\varepsilon}(\eta_1,\eta_2,T) \, ,
}
with
\eq{\label{WWTpr}
  L^{\,\varepsilon}_2(\eta_1,\eta_2) &=\alpha \left(  {1\over
        30} \eta_1\,\partial^3 \eta_2  -{1\over
        30} \partial^3\eta_1\,\eta_2
+{1\over    20} \partial^2\eta_1\,\partial\eta_2-{1\over
        20} \partial\eta_1\,\partial^2\eta_2\right)\\
    &\phantom{=}-{3\hbar\beta\over 10}  \left(  {1\over
        2} \eta_1\,\partial^3 \eta_2  -{1\over
        2} \partial^3\eta_1\,\eta_2
-{1\over    2} \partial^2\eta_1\,\partial\eta_2+{1\over
        2} \partial\eta_1\,\partial^2\eta_2\right)\, , \\[0.3cm]
         L^{\,\varepsilon}_3(\eta_1,\eta_2,T) &=\beta\, \left(
     \eta_1 \partial\eta_2-\partial \eta_1\, \eta_2\right)\,
    T\,.
}
Please note the explicit first order quantum correction in $
L^{\,\varepsilon}_2(\eta_1,\eta_2)$
and the infinitely many quantum corrections hidden in the $\hbar$
dependence of $\beta$.

\subsection{Quantum L$_\infty$ relations with two symmetry parameters} 
\label{checkingrelations}

Having determined the quantum corrected $L_n$ products for the ${\cal W}_3$ algebra, let us now state
 and check the quantum L$_\infty$ relations 
\eq{  \label{nochmalkorrekturen}
          {\cal J}^{\rm qu}_{m+n}(\epsilon_1, \dots , \epsilon_m, x_1,\ldots, x_n)+\sum_{\substack{ (y_1,\ldots, y_k) \\ \rightarrow (x_1, \dots, x_n) }}  \hbar^\xi \; {\cal
            J}^{\rm qu}_{m+k}(\epsilon_1, \dots,  \epsilon_m, \underbrace{y_1,\ldots, y_k}_{\to (x_1,\ldots,x_n)})=0
}
when plugging in exactly two symmetry  parameters. These are the ones that are equivalent to the quantum closure condition \eqref{closureagain}.

\subsubsection*{Quantum corrections to the L$_\infty$ relations}

The distinguished new feature of the definition of quantum L$_\infty$
algebras is the second term in \eqref{nochmalkorrekturen} where the
contractions appear. Let us therefore first list the L$_\infty$
relations that are non-trivially corrected by such contraction
terms.

Since we plug in two symmetry  parameters and we need at least two fields
to be able to contract, we must have at least four inputs in
\eqref{nochmalkorrekturen}. But since the highest $L_n$ product is
$L_3$, all relations ${\cal J}^{\rm qu}_6 , {\cal J}^{\rm qu}_7, \dots
= 0$ are automatically satisfied. To
further trivialize most cases we can use that the only non-trivial
$L_3$ products are $L_3^W(\eta, T,T)$ and $L_3^W (\eta_1, \eta_2,
T)$. Since the first $L_3$ always maps into the kernel of the second
$L_3$, for $ {\cal J}^{\rm qu}_5\sim L_3 L_3$ one can conclude 
\eq{
{\cal J}^{\rm qu}_5 (\epsilon_i, \epsilon_j, \mathbf W, \mathbf W , \mathbf W ) = 0 \, .
}
In a similar vein, evaluating  \eqref{ininftyrel2} one finds that trivially
\eq{
{\cal J}^{\rm qu}_4 (\boldsymbol{\varepsilon}, \boldsymbol{\varepsilon}, W, W) &= 0 \, , \\
{\cal J}^{\rm qu}_4 (\varepsilon_1, \varepsilon_2, \mathbf W, \mathbf W ) &= 0 \, , \\
{\cal J}^{\rm qu}_4 ( \varepsilon, \eta, W, T) &= 0
 \, . 
}
The only non-zero contraction terms can  therefore arise in the terms
\eq{
{\cal J}^{\rm qu}_4 (\epsilon, \eta ,  \contraction{}{ T}{, }{ T}  T,  T ) \, , \quad 
{\cal J}^{\rm qu}_4 (\eta_1, \eta_2 ,  \contraction{}{W}{ ,}{ T}  W,  T   )  \, , \quad 
{\cal J}^{\rm qu}_4 (\eta_1, \eta_2 ,   \contraction{}{ T}{ ,}{ T}  T,  T  ) \, . 
} 
From the form of the OPEs \eqref{OPES}, one realizes that the contraction
$\contraction{}{ T}{ }{ T}  T  T$ yields terms proportional to $\hbar
T$ and the identity $\hbar \mathbf 1$, while the second contraction
reads $\contraction{}{W}{ }{ T}  W  T \sim \hbar W$. 
Hence the L$_\infty$ relations that are non-trivially corrected by a contraction of a higher L$_\infty$ relation are
\eq{
0&={\cal J}_2^{\rm qu}(\varepsilon,\eta)+\hbar \, {\cal J}^{\rm qu}_4(\varepsilon,\eta,\underbrace{T,T}_{\rightarrow \mathbf{1}}) \, , \\
0&={\cal J}_3^{\rm qu}(\varepsilon,\eta,T)+\hbar \, {\cal J}^{\rm qu}_4(\varepsilon,\eta,\underbrace{T,T}_{\rightarrow T}) \, ,  \\
0&={\cal J}_2^{\rm  qu}(\eta_1,\eta_2)+\hbar \, {\cal J}_4^{\rm qu}(\eta_1,\eta_2,\underbrace{T,T}_{\rightarrow \mathbf{1}}) \, , \\
0&={\cal J}_3^{\rm qu}(\eta_1,\eta_2,T)+\hbar \, {\cal J}_4^{\rm qu}(\eta_1,\eta_2,\underbrace{T,T}_{\rightarrow T}) \, , \\
0&={\cal J}_3^{\rm qu}(\eta_1,\eta_2,W)+\hbar \, {\cal J}^{\rm qu}_4(\eta_1,\eta_2,\underbrace{T,W}_{\rightarrow W})\,.
}

Following  the logic of section \ref{quentumlinfty}, we  will now explicitly evaluate the contractions appearing  in
these quantum L$_\infty$ relations. We start with terms arising  from
contractions of the L$_\infty$ relation  ${\cal J}^{\rm qu}_4 (\eta_1,
\eta_2 , T,T  ) $. In a first step we find
\eq{\label{VfB}
{\cal J}^{\rm qu}_4(\eta_1,\eta_2,T,T)=&-L^{T}_2\big( \,L^{\,\varepsilon}_3(\eta_1,\eta_2,T),T\, \big)    
      \\ &+\tfrac{1 }{ 2}L^{T}_2 \big( \,\eta_2,L^{W}_3(\eta_1,T,T)  \big )
     - \tfrac{1 }{ 2} L^{T}_2\big( \,\eta_1,L^{W}_3(\eta_2,T,T) \, \big) \, . 
     }
Recall that every L$_\infty$ relation collects the contribution of the
form $(\eta_1 \eta_2) \, (T \star T)$. While the terms in the second
line are already of this form, the first term is not, so that 
the non-associativity of the $\star$-product \eqref{nonassociativity} 
is expected to induce contractions.
Inserting the explicit expression of the $L_n$
products into the first term 
yields
\eq{\label{l23} \,
-L^{T}_2\big( \,L^{\,\varepsilon}_3(\eta_1,\eta_2,T),T\, \big)=-2\, \beta \, \big( \,\partial(f T)\star T\, \big)-\beta\, \big( \,(f  T)\star \partial T\, \big)\, , 
}
where we abbreviated $f:= \eta_1 \,  \partial\eta_2-\partial \eta_1 \,  \eta_2$. 
Using the normal ordering prescription \eqref{NOproduct} and its
function linearity in the second argument we find for the first term
\begin{align}
-2 \, \beta \, &\Big( \,\partial(f T)\star T \, \Big)(z) \nonumber \\&=-\beta\,\Bigg(\oint {dy \over 2 \pi i} {f(y)\, \, T(y)\circ T(z)\over (y-z)^2}+\partial f(z)\, N(TT)(z)+f(z) \, N(T\, \partial T)\Bigg) \nonumber \\
&=-\beta  \, \Bigg({c \hbar \over 240} \, \partial^5 f(z)+{\hbar \over 3}\, \partial^3 f(z) \,  T(z)+{\hbar \over 2} \, \partial^2 f(z) \,  \partial T(z)\\
&\phantom{=====}+2 \, \partial f(z) \, N(TT)(z)+f(z) \, \partial N(TT)(z)\Bigg) \nonumber \, . 
\end{align} 
Evaluating the second term in \eqref{l23} similarly gives
\begin{align} 
-\beta   & \Big( (f T)\star \partial T\Big)(z)=   \\
 &\phantom{==} -{\beta \over 2}\bigg({c\hbar\over 60}\partial^5 f(z)+{2 \hbar \over 3}\partial^3 f(z)\, T(z)+{3\hbar\over 2}\partial^2 f(z)\,\partial T(z) +f(z)\,\partial N(TT)\bigg) \, . \nonumber
\end{align}
Putting both terms together results in
\eq{
-L^{T}_2\big( \,L^{\,\varepsilon}_3(\eta_1,&\eta_2, T),T\, \big)=-{\beta \hbar c \over 80}\, \partial^5 f(z) \\[0.1cm] 
&-{\beta \hbar \over 3}\, \partial^3 f(z) \, T(z)-{5 \beta \hbar \over 4} \, \partial^2 f(z)\, \partial T(z)
-{\beta \hbar\over 2} \, \partial f(z) \, \partial^2 T(z) \\[0.1cm]
&-2\, \beta\,  \partial f(z)\, N(TT)(z)-{3\beta\over 2} \, f(z)\, \partial N(TT)(z) \,  
}
so that we can directly read off
\begin{align}
\label{correctionsetaetaTT}
\hbar \, {\cal J}_4^{\rm qu}(\eta_1,\eta_2,\underbrace{T,T}_{\rightarrow \mathbf{1}})&=-{\beta \hbar c \over 80}\partial^5 f(z) \, , \\
\hbar \, {\cal J}_4^{\rm qu}(\eta_1,\eta_2,\underbrace{T,T}_{\rightarrow T})&=-{\beta \hbar \over 3}\partial^3 f(z)\,T(z)-{5 \beta \hbar \over 4} \partial^2 f(z)\,\partial T(z) -{\beta \hbar\over 2} \partial f(z)\,\partial^2 T(z)\, . \nonumber
\end{align} 
Computing the other contractions is more lengthy, but follows the same steps. Let us therefore only state the results
\begin{align} \label{correctionsepsetaTT}
\hbar \,{\cal J}^{\rm qu}_4(\varepsilon,\eta,\underbrace{T,T}_{\rightarrow T})=&-{4\hbar \beta \over 3} \, \left (\partial \eta \, \partial^3 \varepsilon-\tfrac{1}{2}\,  \eta \, \partial^4 \varepsilon \right)\, T
-2\hbar \beta \, (\partial \eta \,   \partial^2 \varepsilon-\tfrac{1}{
  3} \, \eta  \, \partial^3\varepsilon)\,\partial T \nonumber  \\[-0.1cm] 
&\qquad -\hbar \beta\,\eta \, \partial^2 \varepsilon \,\partial^2 T \, , \\[0.3cm]
\hbar \,{\cal J}^{\rm qu}_4(\varepsilon,\eta,\underbrace{T,T}_{\rightarrow \mathbf{1}})=&-{\beta \hbar c \over 40}\, (\partial \eta \, \partial^5\varepsilon+\tfrac{1}{2}\eta \, \partial^6\varepsilon)  \, , \nonumber 
\end{align}
and finally
\eq{ \label{correctionsetaetaWW}
\hbar\, {\cal J}_4(\eta_1,\eta_2,\underbrace{T,W}_{\rightarrow W})=&-\,\frac{3\beta \hbar }{ 4}(\partial \eta_1 \, \partial^2  \eta_2-\partial\eta_2 \, \partial^2 \eta_1)\, \partial W\\
& -{3\beta \hbar \over 2}\, \partial^2 f\,\partial W -{9\beta \hbar \over 4} \,  \partial f \, \partial^2 W \, . 
}

\subsubsection*{Checking the quantum L$_\infty$ relations}

We are now in the position to  state and check the quantum L$_\infty$
relation with two symmetry  parameters. 
We will sort them according to their appearance in the quantum closure condition \eqref{closureagain} with $i,j,k \in \{T, W\}$. 

\begin{itemize}
\item \textbf{(TT,T):} The closure condition \eqref{closureagain} with $(ij,k) = (TT,T)$ is equivalent to
\eq{
0 &= {\cal J}^{\rm qu}_2(\varepsilon_1,\varepsilon_2) \\[0.1cm]
&= -L^{T}_1 \big
(L^{\,\varepsilon}_2(\varepsilon_1,\varepsilon_2)\, \big)+ L^{T}_2\big( \,L^{T}_1(\varepsilon_1),
                 \varepsilon_2\, \big) + L^{T}_2\big( \,\varepsilon_1,L^{T}_1(\varepsilon_2)\, \big)
}
and
\eq{
0 &= {\cal J}^{\rm qu}_3(\varepsilon_1,\varepsilon_2,T) \\[0.1cm]
&= L^{T}_2\big( \,L^{\,\varepsilon}_2(\varepsilon_1,\varepsilon_2),T\, \big)+L^{T}_2\big( \,L^{T}_2(\varepsilon_2,T),
                 \varepsilon_1\, \big) + L^{T}_2\big( \,L^{T}_2(T,\varepsilon_1),\varepsilon_2\, \big) \, .
}
Inserting \eqref{TT,T} these relations are readily  checked to be satisfied.

\item \textbf{(TT,W):} There is only one non-trivial relation
\begin{align}
0&={\cal J}^{\rm qu}_3(\varepsilon_1,\varepsilon_2,W)\\[0.1cm]
&=L^{W}_2\big( \,L^{\,\varepsilon}_2(\varepsilon_1,\varepsilon_2),W\, \big)+L^{W}_2\big( \,L^{W}_2(\varepsilon_2,W),
                 \varepsilon_1\, \big) + L^{W}_2\big( \,L^{W}_2(W,\varepsilon_1),\varepsilon_2\, \big) \, , \nonumber
\end{align}
that is also directly satisfied.

\item \textbf{(TW,T):} One finds the single non-trivial relation
\eq{
0&={\cal J}^{\rm qu} _3(\varepsilon,\eta,W)\\
  &=  L^{T}_2\big( \,L^{\,\eta}_2(\varepsilon,\eta),W\, \big)+
 L^{T}_2\big( \,L^{T}_2(\eta ,W),\varepsilon\, \big) +
L^{T}_2\big( \,L^{W}_2(W,\varepsilon),\eta\, \big) \,.
}
As before, a short computation shows that this equation is satisfied
without any constraints.

\item \textbf{(TW,W):} This is the first truly interesting case, as  the
  closure condition involves a contribution from a contraction  
\eq{
0&={\cal J}^{\rm qu}_2(\varepsilon,\eta)+\hbar \, {\cal J}^{\rm qu}_4(\varepsilon,\eta,\underbrace{T,T}_{\rightarrow \mathbf{1}}) \, , \\
0&={\cal J}^{\rm qu}_3(\varepsilon,\eta,T)+\hbar \, {\cal J}^{\rm qu}_4(\varepsilon,\eta,\underbrace{T,T}_{\rightarrow T}) \, , \\
0&={\cal J}^{\rm qu}_4(\varepsilon,\eta,T,T)\,.
}
When evaluating these relations, the contraction terms computed in
\eqref{correctionsepsetaTT} are crucial. Like in the classical case,
the first equation is satisfied for $\alpha=2$. Note that terms from
the quantum part of $L_2(\eta,T)$ get exactly canceled by the
quantum correction from the contraction. The second equation is indeed satisfied for
$\beta={16\alpha \over 5c+22\hbar}$, the value of the quantum ${\cal
  W}_3$ algebra.
The third relation  holds without giving any constraints 
on $\alpha, \beta$.

\item \textbf{(WW,T):} 
In this case the closure is equivalent to the quantum L$_\infty$ relations
\eq{\label{qmWWt}
0&={\cal J}^{\rm qu}_2(\eta_1,\eta_2)+\hbar \, {\cal J}_4^{\rm qu}(\eta_1,\eta_2,\underbrace{T,T}_{\rightarrow \mathbf{1}}) \, , \\
0&={\cal J}_3^{\rm qu}(\eta_1,\eta_2,T)+\hbar \, {\cal J}_4^{\rm qu}(\eta_1,\eta_2,\underbrace{T,T}_{\rightarrow T}) \, , \\
0&={\cal J}_4^{\rm qu}(\eta_1,\eta_2,T,T)\,.
}
Again, the contraction terms \eqref{correctionsetaetaTT} are
needed. The first equation is satisfied for $\alpha=2$ and the second
for $\beta={16 \alpha \over 5c+22\hbar}$. Again, the quantum corrected L$_\infty$ relations fix the open constants exactly to the values expected for the quantum ${\cal W}_3$ algebra. The third equation holds independently of the numerical values of $\alpha, \beta$.

\item \textbf{(WW,W):} The quantum L$_\infty$ relations  equivalent to closure are
\eq{
0&=  {\cal J}^{\rm qu}_3(\eta_1,\eta_2,W)+\hbar \, {\cal J}^{\rm qu}_4(\eta_1,\eta_2,\underbrace{T,W}_{\rightarrow W}) \, , \\
0&={\cal J}^{\rm qu}_4(\eta_1,\eta_2,T,W)\,.
}
After inserting the contraction term \eqref{correctionsetaetaWW},
both equations hold independent of $\alpha$ and $\beta$.

\end{itemize}

\subsection{L$_\infty$ relations with three symmetry  parameters}

After we have checked the L$_\infty$ relations with two symmetry 
parameters, it remains to evaluate those with three symmetry  parameters.
Recall  that these are equivalent to the Jacobi identity
\eq{ \label{stuttgarterkickers}
\sum_{\rm cycl} \big[ \delta^{\rm qu}_{\varepsilon_i}, [\delta^{\rm qu}_{\varepsilon_j},\delta^{\rm qu}_{\varepsilon_k}]\big]=0\,.
}
For three symmetry  parameter insertions,  ${\cal J}_n=0$ is trivially
satisfied for   $n\ge 5$  in the case of the ${\cal W}_3$
algebra. Therefore,  there cannot be any correction terms arising
from contractions.
Again sorting them according to the triplet $(ijk)$ in \eqref{stuttgarterkickers},
the quantum L$_\infty$ relations read as follows:
\begin{itemize}
\item {\bf (TTT):}
\eq{
0=L^{\,\varepsilon}_2\big( \,L^{\,\varepsilon}_2(\varepsilon_1,\varepsilon_2),\varepsilon_3\, \big)+L^{\,\varepsilon}_2\big( \,L^{\,\varepsilon}_2(\varepsilon_3,\varepsilon_1),\varepsilon_2\, \big)+L^{\,\varepsilon}_2\big( \,L^{\,\varepsilon}_2(\varepsilon_2,\varepsilon_3),\varepsilon_1\, \big) \,.\nonumber
}

\item {\bf (TTW):}
\eq{
0=L^{\,\eta}_2\big( \,L_2^{\,\varepsilon}(\varepsilon_1,\varepsilon_2),\eta\, \big)+L^{\,\eta}_2\big( \,L_2^{\,\eta}(\eta,\varepsilon_1),\varepsilon_2\, \big)+L^{\,\eta}_2\big( \,L_2^{\,\eta}(\varepsilon_2,\eta),\varepsilon_1\, \big) \,.\nonumber
}

\item {\bf (WWT):}
\begin{align*}
&0=L^{\,\varepsilon}_2\big( \,L_2^{\,\varepsilon}(\eta_1,\eta_2),\varepsilon\, \big)+L^{\,\varepsilon}_2\big( \,L_2^{\,\eta}(\varepsilon,\eta_1),\eta_2\, \big)+L^{\,\varepsilon}_2\big( \,L_2^{\,\eta}(\eta_2,\varepsilon),\eta_1\, \big)\\[0.1cm]
	&\phantom{0=}+L^{\,\varepsilon}_3\big( \,\eta_1,\eta_2,L^{T}_1(\varepsilon)\, \big)\,,
\\[0.3cm]
& 0=-L^{\,\varepsilon}_2\big( \,L^{\,\varepsilon}_3(\eta_1,\eta_2,T),\varepsilon\, \big)+L^{\,\varepsilon}_3\big( \,L^{\,\eta}_2(\eta_1,\varepsilon),\eta_2,T\, \big)\,,\\[0.1cm]
& \phantom{0=} -L^{\,\varepsilon}_3\big( \,L^{\,\eta}_2(\eta_2,\varepsilon),\eta_1,T\, \big)+L^{\,\varepsilon}_3\big( \,L^{T}_2(T,\varepsilon),\eta_1,\eta_2\, \big) \,.
\end{align*}
The first ${\cal J}_3$-type relation requires $\beta={16\alpha \over 5c+22\hbar}$
to hold  and, due to the appearance of the non-vanishing last term,
features  that the two-product $L_2$ violates its  Jacobi identity.

\item {\bf (WWW):}
\begin{align*}
0&=L^{\,\eta}_2\big( \,L_2^{\,\varepsilon}(\eta_1,\eta_2),\eta_3\, \big)+L^{\,\eta}_2\big( \,L_2^{\,\varepsilon}(\eta_3,\eta_1),\eta_2\, \big)+L^{\,\eta}_2\big( \,L_2^{\,\varepsilon}(\eta_2,\eta_3),\eta_1\, \big) \,,  \\[0.2cm]
0&=L^{\,\eta}_2\big( \,L_3^{\,\varepsilon}(\eta_1,\eta_2,T),\eta_3\, \big)+L^{\,\eta}_2\big( \,L_3^{\,\varepsilon}(\eta_3,\eta_1,T),\eta_2\, \big)+L^{\,\eta}_2\big( \,L_3^{\,\varepsilon}(\eta_2,\eta_3,T),\eta_1\, \big) \, , \\[0.2cm] 
0&=L^{\,\varepsilon}_3\big( \,L_2^{T}(\eta_1,W),\eta_2,\eta_3\, \big)+L^{\,\varepsilon}_3\big( \,L_2^{T}(\eta_2,W),\eta_3,\eta_1\, \big) \\& \qquad +L^{\,\varepsilon}_3\big( \,L_2^{T}(\eta_3,W),\eta_1,\eta_2\, \big)\,.
\end{align*}
\end{itemize}

\section{Summary and Conclusions}

This completes the proof that the quantum ${\cal W}_3$ algebra is
an example for a quantum L$_\infty$
algebra as defined in section \ref{quentumlinfty}. 
Like for the classical  ${\cal W}_3$ algebra, the quantum corrected
relations with two inputs gave the constraint $\alpha=2$ and the
relations with three inputs ${\cal J}^{\rm qu}_3=0$ required
$\beta={16 \alpha \over 5c +22\hbar}$. The only other non-trivial
higher order relations were satisfied without any further
constraint. The L$_\infty$ relations with three symmetry parameters were
essentially  the same as in the classical case.

Let us emphasize that the quantum contractions  in
\eqref{quantumrelas} are necessary for the L$_\infty$ relations to hold.
This means that the quantum ${\cal W}_3$ algebra does neither define a
classical nor a loop L$_\infty$ algebra (as appeared for CSFT),  but this new type of a quantum L$_\infty$ algebra. 
Of course the higher products in CSFT and for quantum ${\cal W}$
algebras are
different from the onset. In the latter case they involve the non-associative normal ordered
product of 2d CFT, whereas in the former case  they are the loop
corrected n-vertices of CSFT. 
Thus, it seems that  for global and gauge symmetries
there does not exist a
unique  version of a physically reasonable definition of an  L$_\infty$
algebra for a quantum theory. 

We expect that in general the whole class of ${\cal W}$ algebras
yields further examples for quantum L$_\infty$ algebras, since all of
them have a closing symmetry  algebra that involves normal ordered
products as defined in CFT. As in the classical case, also higher
$n$-products will be non-trivial. Since our analysis of quantum {\cal
  W}-algebras is restricted to non-trivial elements in $X_0 \oplus
X_{-1}$, it is not obvious whether and how this structure generalizes
to more general gradings.


\vskip3em
\noindent
\emph{Acknowledgments:} We are grateful to Andreas Deser for
discussions.




\clearpage
\bibliography{references2}  
\bibliographystyle{utphys}


\end{document}